\documentclass{article}
\usepackage[utf8]{inputenc}
\usepackage[T1]{fontenc}
\usepackage{times} 
\usepackage[margin=1.05in]{geometry}
\usepackage{graphicx}
\usepackage{amsmath,amssymb}
\usepackage{booktabs}
\usepackage{algorithm}
\usepackage{algpseudocode}
\usepackage[numbers]{natbib}
\usepackage[hidelinks]{hyperref}
\usepackage{float}
\usepackage{xcolor}

\newfloat{algorithm}{tbp}{alg}
\floatname{algorithm}{Algorithm}

\title{\textbf{Duration-Aware World Models for Sleep Staging\\[1ex]
with Dwell-Time Constraints}}
\author{Junhui Gao\\[2ex]
Shenzhen Peng Brain Technology Co., Ltd.\\[2ex]
\href{mailto:jhgao68@163.com}{jhgao68@163.com}}
\date{}

\newcommand{\datasetA}{SHHS-1}
\newcommand{\datasetB}{Sleep-EDF}

\begin{document}
\maketitle

\begin{abstract}
Closed-loop neuromodulation requires sleep staging that respects
stage-specific minimum dwells: stimulation policies switch by sleep stage, and
a policy change is permitted only after the current stage has persisted for its
minimum duration. Discriminative classifiers fragment stage trajectories---for
example, a 30 s Wake segment inside a 10 min N2 run---and even hidden
semi-Markov smoothing leaves 11.3\% of segments below the minimum dwell
threshold. This paper presents a world model in which a recurrent state-space model
(RSSM) learns dwell structure endogenously from EEG, via duration-aware latent
sampling, a survival-curve dwell prior, and transition-matrix calibration, and
a trajectory-level hard-constraint layer guarantees the minimum dwell at
decoding. On \datasetA{} (200 subjects, 160/20/20, three repeated splits), the
model reaches a test accuracy of $0.690 \pm 0.053$ (95\% CI over 3 folds)
versus $0.641 \pm 0.042$ for an identically split discriminative baseline, with
macro-F1 $+17.5$ pp and better accuracy on all three folds. On \datasetB{} the
constraint layer cuts below-threshold segments from 71.8\% to 0.1\% with no
accuracy loss ($0.734 \to 0.737$), and an online locked scheduler reduces RL
policy switches by 87\% per night while keeping every policy run at least 90 s.
To the best of the author's knowledge, this is the first system to couple
learned dwell statistics with a decoding-time hard guarantee. Code and pretrained weights
will be released.
\end{abstract}

\section{Introduction}
\label{sec:intro}

Real-time closed-loop neuromodulation systems modulate stimulation in response
to the instantaneous sleep stage: different stages call for different
stimulation policies, and an RL policy selected for a given stage is trained
for that stage's dynamics. The stage sequence produced by the sensing
subsystem therefore acts as a switching signal. A key requirement in
this setting is \emph{dwell-time stability}: each stage must persist for a
minimum duration before the system is allowed to switch policies; otherwise
the controller chatters between policies on noisy boundaries. In practice,
stage-specific minima differ: N2 is expected to last minutes, whereas N1 and
REM transitions are shorter, and the operator needs a \emph{guarantee} that
no stage run falls below its minimum.

Current sleep staging systems are discriminative: they map each 30 s EEG epoch
to a stage. State-of-the-art networks such as U-Time \citep{perslev2019utime},
DeepSleepNet \citep{supratak2017deepsleepnet} and SleepTransformer
\citep{phan2022sleeptransformer} achieve high epoch-wise accuracy, yet their
output trajectories are fragmented, containing many 1--2 epoch runs that
violate physiological and application-level dwell constraints. Post-processing
with hidden Markov / semi-Markov models
\citep{rabiner1989tutorial,flexer2005reliable} smooths the sequence without
hard guarantees: it reduces violations but cannot eliminate them. The discriminative
paradigm models each epoch in isolation (or with local context) and has no
mechanism to represent or enforce \emph{how long a stage should persist}.

Sleep is organized on two time scales. On the \emph{epoch} scale (30 s), each
window carries the physiological features that identify a stage, and
discriminative networks excel there. On the \emph{structural} scale (minutes
to hours), stages persist for characteristic durations and recur in cycles; it
is on this scale that the application-level requirement---a minimum dwell per
stage---is defined. Discriminative staging models only the first scale and
delegates the second to post-processing. The mismatch matters in closed-loop
neuromodulation: switching a stimulation policy is an action taken on the
structural scale, so the quantity that matters to the controller is the
\emph{run-level} structure of the stage sequence, not the epoch-level
accuracy of its labels.

This paper instead builds on \emph{world models}
\citep{ha2018worldmodels,hafner2019plaNet,hafner2020dreamer,hafner2021dreamerv2}.
A world model learns a latent dynamical system over the data; given a sleep
night, it maintains a latent state that evolves over time and from which the
stage and the EEG features can be reconstructed. Because the latent state is
\emph{generative} and \emph{temporal}, it is well suited to representing
dwell structure. A recurrent state-space model (RSSM) is extended with a
duration-aware latent sampler: an explicit dwell counter and a
continue/switch Bernoulli variable that decides whether the current stage
persists or transitions. On top of this, two empirically fitted priors are added---the survival-curve
dwell prior and the transition-matrix calibration---and a trajectory-level
hard-constraint layer guarantees that every generated run is at least $K[s]$
epochs for stage $s$.

This paper makes three contributions.
\begin{itemize}
\item \textbf{A generative world model with endogenous dwell structure.} The
paper shows that a recurrent state-space model (RSSM)
trained on EEG can stage sleep while learning dwell structure endogenously,
and that under identical data splits it \emph{outperforms} a discriminative
baseline (test accuracy $0.690 \pm 0.053$ vs $0.641 \pm 0.042$; macro-F1
$+17.5$ pp; better on all 3 repeated splits).
\item \textbf{Three progressive mechanisms and a hard constraint layer.}
Duration-aware latent sampling, a survival-curve dwell prior, and transition
calibration bring the dreamed dwell distribution close to physiological
statistics without loss of accuracy (val $0.714 \to 0.723$);
a trajectory-level hard constraint (latent forcing plus run merging) drives
below-minimum-dwell segments from 71.8\% to 0.1\% on \datasetB{} while
accuracy improves from $0.734$ to $0.737$.
\item \textbf{Causal online scheduling.} An online locked scheduler reduces
reinforcement-learning policy switches by 87\% per night with every run at
least 90 s, showing that the guarantee carries over from batch staging to a causal
online controller.
\end{itemize}

\section{Related Work}
\label{sec:related}

\paragraph{Discriminative sleep staging.}
Automatic sleep staging from single-channel EEG has been approached with
convolutional networks \citep{supratak2017deepsleepnet,sors2018cnn,chambon2018deep},
sequence models \citep{phan2022sleeptransformer}, and fully convolutional
segmentation networks \citep{perslev2019utime,perslev2021usleep}. These models are trained with
per-epoch cross-entropy; temporal consistency is either ignored or handled by
post-hoc smoothing (HMM/HSMM) \citep{rabiner1989tutorial,flexer2005reliable}.
None of them learns a generative prior over stage runs, and none provides a
guarantee on minimum dwell.

\paragraph{World models.}
World models learn predictive latent representations of environments
\citep{ha2018worldmodels}. The recurrent state-space model (RSSM) of
PlaNet/Dreamer \citep{hafner2019plaNet,hafner2020dreamer} and its discrete
successors \citep{hafner2021dreamerv2,hafner2023dreamerv3} learn a stochastic
latent dynamics and train with KL balancing and latent imagination.
Transformer world models \citep{robine2023twm} replace the recurrent core with
a transformer. The use of an RSSM for EEG sleep staging has not, to the best of the
author's knowledge, been reported; the closest analogies are recent works using latent dynamics for
physiological time series, which remain rare. Score-based generative staging \citep{vangorp2024generative} instead samples labels directly from data and, unlike the RSSM, offers no forward rollout in latent space.

\paragraph{Dwell constraints and sequence post-processing.}
HMM/HSMM approaches impose explicit state-duration models and can encode
minimum-dwell as part of the duration distribution \citep{rabiner1989tutorial,yu2010hsmm}.
However, the guarantee is probabilistic, and the duration model is a
post-hoc prior decoupled from the feature extractor. Here the dwell structure
is learned inside the latent dynamics, and the minimum is enforced as a hard
constraint at decoding time; the constraint layer itself is model-agnostic:
it applies to the output of any staging system, as demonstrated on
discriminative outputs in Section~\ref{sec:constraint-exp}.

\section{Method}
\label{sec:method}

\subsection{Problem setting}
\label{sec:setting}

A night is a sequence of 30 s EEG epochs. The input window is denoted
$x_t \in \mathbb{R}^{1 \times 3000}$ (single channel, 100 Hz) and the stage
label $y_t \in \{0,\dots,4\}$ (Wake, N1, N2, N3, REM) following the
Rechtschaffen--Kales and AASM conventions
\citep{rechtschaffen1968manual,berry2017aasm}. A \emph{run} (dwell) is
a maximal contiguous block of identical stages; its length is the dwell
$L \in \mathbb{Z}_{\ge 1}$ (in epochs). The application constraint is a vector
$K \in \mathbb{Z}_{\ge 1}^5$: every run of stage $s$ must satisfy $L \ge K[s]$.
In this paper $K = \{2,2,6,4,4\}$ for Wake, N1, N2, N3, REM (30 s per epoch).

\subsection{RSSM backbone}
\label{sec:rssm}

\begin{figure}[t]
\centering
\includegraphics[width=0.95\textwidth]{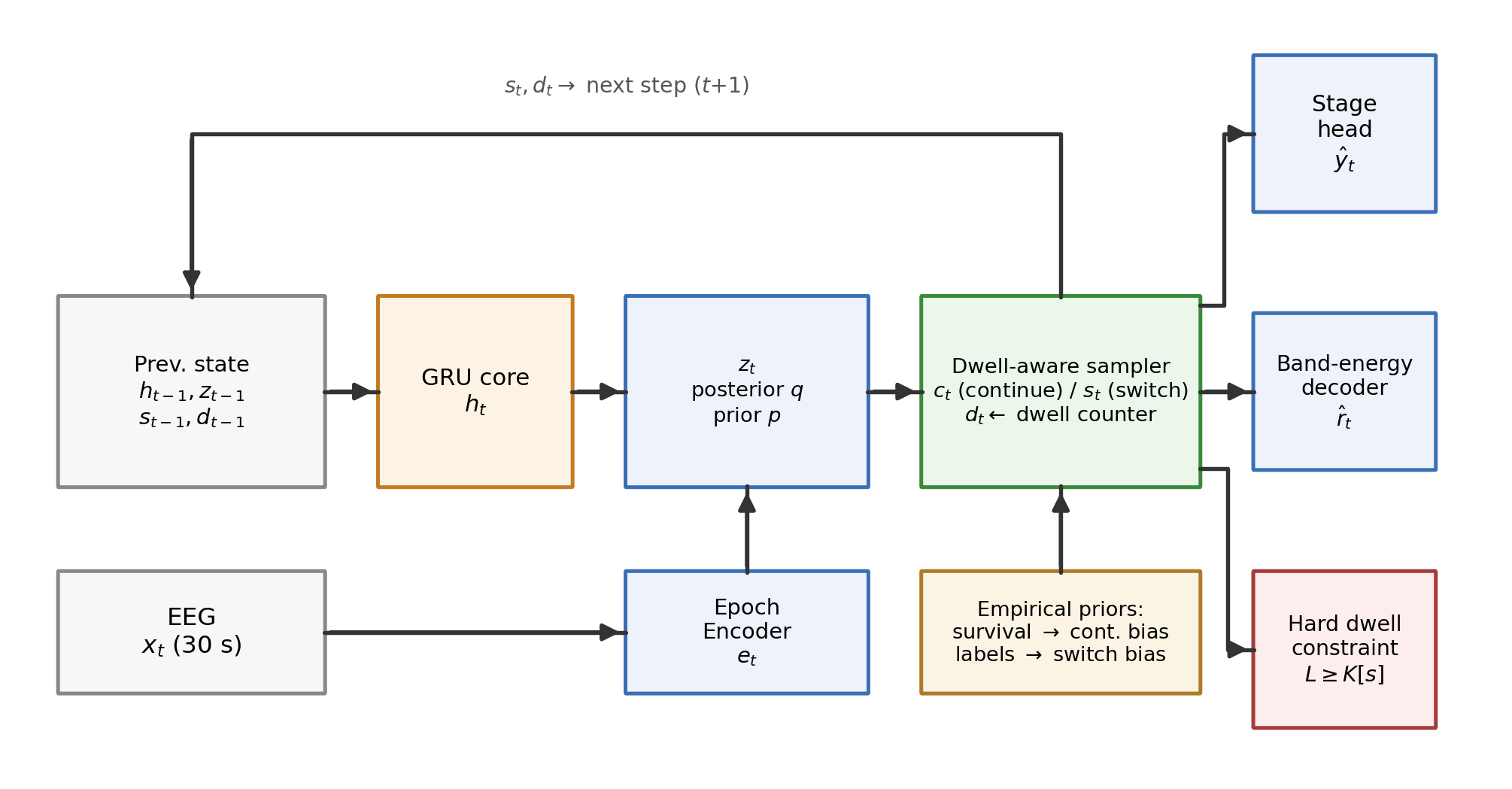}
\caption{Duration-aware RSSM world model. An epoch encoder maps each 30 s EEG
window to an embedding; a GRU core evolves the latent state; the
dwell-aware sampler decides continue/switch with a dwell counter $d_t$,
guided by empirically fitted priors; the stage head and band-energy decoder
read out the stage and the EEG spectrum. A hard dwell constraint guarantees
every run is at least $K[s]$ epochs. Final calibrated model: val/test accuracy 0.723/0.645 on \datasetA{}.}
\label{fig:arch}
\end{figure}

Figure~\ref{fig:arch} shows the architecture. At a high level, the pipeline
reads each 30 s EEG window through an epoch encoder, evolves a latent state
along a recurrent GRU core, and reads out both the stage and the band-energy
reconstruction from that state; the emitted stage $s_t$ and its dwell counter
$d_t$ are fed back into the next step, so a whole night can be rolled out in
latent space without observations. Three components are new to sleep staging
and are highlighted in Figure~\ref{fig:arch}: the duration-aware sampler that
decides continue/switch, the empirical dwell priors (the survival-curve bias
and the transition calibration), and the trajectory-level hard-constraint
layer; the remaining components follow the standard RSSM design. The RSSM architecture of Hafner
et~al.~\citep{hafner2020dreamer} is adopted. An epoch encoder
(a 1-D convolutional stack) maps $x_t$ to an embedding $e_t$. A GRU core
maintains the recurrent state $h_t$; the stochastic state $z_t$ is sampled
from a Gaussian posterior $q(z_t \mid h_t, e_t)$ during training and from the
prior $p(z_t \mid h_t)$ when predicting. The stage logits are produced by a
classifier over the concatenation $[h_t, z_t, s_t]$ where $s_t$ is the
one-hot stage. The model also reconstructs the EEG frequency-band energies
(delta, theta, alpha, beta) from $[h_t, z_t, s_t]$; the reconstruction loss
anchors the latent state to observable EEG content and prevents posterior
collapse. Training uses KL balancing with free bits ($0.5/0.2/0.3$ for $z$,
continue, and switch), a linear KL warm-up over 15 epochs, and imagination
(when the last $L$ steps of a sequence are predicted from the prior) ramped
from 0 to 20 steps.

\subsection{Duration-aware latent sampling}
\label{sec:duration}

Discriminative classifiers treat stages as independent categorical decisions.
To represent dwell structure in the latent dynamics, the latent state is augmented with a dwell
counter $d_t \in \{0,\dots,D_{\max}\}$ that counts how long the current stage
has persisted, and a Bernoulli \emph{continue} variable $c_t \in \{0,1\}$
that decides whether the current stage persists ($c_t=1$) or a switch is
sampled from the stage-prior logits. At every latent step:
\begin{equation}
\label{eq:step}
\begin{aligned}
&h_t = \mathrm{GRU}\big([h_{t-1}, z_{t-1}, s_{t-1}, e(d_{t-1})]\big),\\
&\mu_p,\sigma_p = p_z(h_t),\quad \mu_q,\sigma_q = q_z(h_t, e_t),\\
&\log p(c_t) = \mathrm{cont}\big([h_t, s_{t-1}, e(d_{t-1})]\big),\\
&\log p(s_t) = \mathrm{switch}\big([h_t]\big),\quad
s_t = c_t\, s_{t-1} + (1-c_t)\, \tilde{s}_t,\\
&d_t = \begin{cases} d_{t-1} + 1 & \text{if } s_t = s_{t-1},\\ 1 & \text{otherwise.}\end{cases}
\end{aligned}
\end{equation}
During training the continue/switch variables are trained with a soft mixture
(the straight-through stage update keeps gradients flowing); during dreaming
they are sampled hard. The classifier and decoder then operate on
$[h_t, z_t, s_t]$ as before. This adds the KL terms $\mathrm{KL}_c$ and
$\mathrm{KL}_t$ for the continue and switch distributions, which are annealed
with the same schedule as $\mathrm{KL}_z$.

\subsection{Empirical dwell prior}
\label{sec:durprior}

\begin{figure}[H]
\centering
\includegraphics[width=0.9\textwidth]{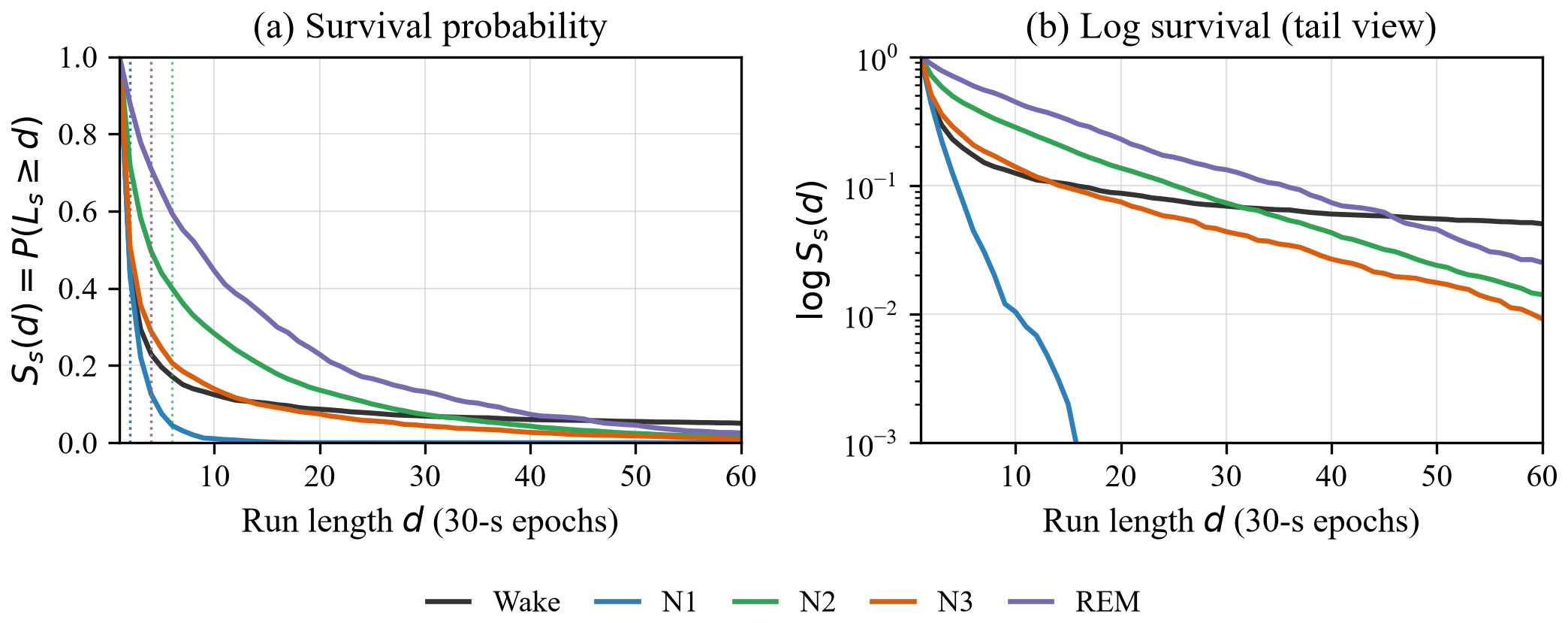}
\caption{Empirical dwell survival curves $S_s(d) = P(L_s \ge d)$ estimated
from the 160 SHHS-1 training subjects; dashed vertical lines mark the
application minima $K_s$. REM dwells are the most persistent
($S_{\mathrm{REM}}(4) = 0.71$), N1 decays fastest ($S_{\mathrm{N1}}(20) \approx 0$),
and N2 carries a heavy tail ($S_{\mathrm{N2}}(10) = 0.28$). The continue-logit
bias in Eq.~(\ref{eq:durbias}) is computed from these curves, which is why the
prior is non-symmetric across stages.}
\label{fig:surv}
\end{figure}

The model learns the dynamics from data, but the dwell distribution is
heavy-tailed and the continue logit is weakly supervised (only through the
switching boundaries). An empirical prior fitted from the
training labels is therefore injected: for each stage $s$, the survival function of real
dwells $S_s(d) = P(L_s \ge d)$ (Figure~\ref{fig:surv}) is computed and the continue logit bias is set to
\begin{equation}
b_s(d) = \log\frac{S_s(d+1)}{S_s(d) - S_s(d+1)},
\qquad d = 1,\dots,D_{\max},
\label{eq:durbias}
\end{equation}
which is the empirical log-odds of continuing given the current dwell. The
bias is added to the continue logit at every latent step
($\log p(c_t) \gets \log p(c_t) + b_{s_{t-1}}(d_{t-1})$). This is a
zero-parameter injection: the network only learns the residual between the
physiology and its own dynamics. A fixed penalty is also applied to the switch
logits for N1 ($-3.0$) because N1 is rare ($\approx 3.1\%$ of epochs) and the
model otherwise over-generates it.

\subsection{Transition-matrix calibration}
\label{sec:transcal}

The switch distribution is also weakly supervised; the empirical
transition-target distribution $\hat{P}(t \mid s)$ is estimated from the training labels
and used to calibrate the switch logits by
\begin{equation}
\label{eq:switchbias}
\mathrm{bias}(s,t) = \log \frac{\hat P(t \mid s)}{1/4},
\end{equation}
i.e., the log ratio to a uniform target. This corrects systematic flow errors that a single scalar penalty cannot
address, such as the over-transfer from N2 to REM, and restores the N3
frequency (N3 segments were being absorbed by REM in the dwell-prior
variant).

\subsection{Trajectory-level hard constraint}
\label{sec:constraint}

Two complementary layers enforce the minimum dwell $K[s]$ at decoding time,
without retraining:
\begin{itemize}
\item \emph{Latent forcing:} during dreaming, if the current stage $s$ has
dwell $d < K[s]$, the continue variable is forced to $c_t = 1$ (no switch is
sampled). This guarantees, inside the latent trajectory, that every run is at
least $K[s]$.
\item \emph{Run merging (observation layer):} for any produced sequence
(argmax or dreamed), every run shorter than $K[s]$ is merged into its
adjacent run (in AASM style, the shorter run is relabeled to the stage of the
longer neighboring run, or to the right neighbor on a tie). This is
model-agnostic and can also be applied to discriminative outputs.
\end{itemize}
The latent forcing is exact for generated trajectories; the merging rule is
exact for any trajectory of finite length and never creates new violations.
Algorithm~\ref{alg:dream} summarizes the dreaming procedure with latent
forcing.

\begin{algorithm}[H]
\caption{Dream with hard dwell constraint}
\label{alg:dream}
\begin{algorithmic}[1]
\Require initial latent state $(h,z,s,d)$; minimum dwells $K[s]$
\Ensure stages $s_1,\ldots,s_T$ with every run at least $K[s]$ epochs
\For{$t = 1 \ldots T$}
  \State update $h_t$, $z_t$ via Eq.~\eqref{eq:step}
  \If{$d_{t-1} < K[s]$}
    \State force $c_t = 1$ \Comment{no switch is sampled}
  \Else
    \State $c_t \sim \mathrm{Bernoulli}(\sigma(\mathrm{cont}))$
  \EndIf
  \If{$c_t = 0$}
    \State $s_t \sim \mathrm{Cat}(\mathrm{switch})$; $d_t = 1$
  \Else
    \State $s_t = s_{t-1}$; $d_t = d_{t-1} + 1$
  \EndIf
\EndFor
\State \Return $s_1, \ldots, s_T$
\end{algorithmic}
\end{algorithm}

\subsection{Training objective}
\label{sec:objective}

Building on the variational auto-encoding objective \citep{kingma2014vae}, the full objective is
\begin{equation}
\label{eq:loss}
L = \mathrm{CE}\big(\mathrm{logits}, y\big) + \lambda_{\mathrm{rec}} \mathrm{MSE}(r, E(x))
+ \beta(t)\Big(\mathrm{KL}_z + \mathrm{KL}_c + \mathrm{KL}_t\Big),
\end{equation}
where the first term is the stage-classification cross-entropy, the second term is the
band-energy reconstruction weighted by $\lambda_{\mathrm{rec}} = 0.3$, $\beta(t)$ is
the annealed KL weight, and the KL terms use free bits. Class imbalance is handled with
inverse-frequency class weights. Training uses Adam ($\mathrm{lr}=10^{-3}$,
weight decay $10^{-4}$), batch size 32, sequence length 100 epochs (50 min,
stride 50), for 60 epochs with early stopping on validation accuracy.

\section{Experiments}
\label{sec:exp}

\subsection{Data and preprocessing}
\label{sec:data}

\datasetA{} \citep{quan1997shhs} is the first cohort of the Sleep Heart
Health Study; the full cohort contains 5{,}793 recordings, of which the
first 1{,}000 were converted locally (953 pass quality checks), and the
first 200 subjects by ID are taken for this study. EEG (C3-A2, 125 Hz) is
downsampled to 100 Hz by FFT resampling and z-scored per 30 s epoch.
Labels follow five classes (Wake, N1, N2, N3, REM); stage distribution is
highly imbalanced (N1 $\approx 3.1\%$). Subject-level splits: 160 train / 20
val / 20 test. For statistical rigor, the split is repeated 3 times with
different random permutations (seeds 0--2), and means $\pm$ 95\% CI are
reported (paired across models on the same folds). For cross-dataset evaluation of the
constraint layer, \datasetB{} is used (Expanded FPz-Cz, 32 held-out test nights)
\citep{kemp2000sleepedf,goldberger2000physionet}.

\subsection{Implementation}
\label{sec:impl}

All world-model variants share the training protocol of
Section~\ref{sec:objective}: sequence length 100, stride 50, batch 32, 60
epochs, warm-up 15, $\lambda_{\mathrm{rec}}=0.3$, free bits $0.5/0.2/0.3$, and
imagination ramped to 20. The discriminative baseline is a 1-D convolutional
classifier (the same encoder as the world model plus a linear head) trained
from scratch under the identical split and 60-epoch budget with
sqrt-inverse-frequency class weights. Both models run on a single NVIDIA RTX
4060 Laptop GPU.

\subsection{Training health and accuracy ablation}
\label{sec:ablation}

Table~\ref{tab:ablation} reports the four-variant ablation. All three KL terms stay
in a healthy regime ($\mathrm{KL}_z \in [1.7,2.0]$, $\mathrm{KL}_c \in
[0.15,0.19]$, $\mathrm{KL}_t \in [0.27,0.30]$, reconstruction $\approx 0.25$),
in contrast to the collapsed training observed when the same architecture
was trained on only 47 subjects of \datasetB{} (not shown). The dwell mechanisms
do not degrade accuracy; the final calibrated model reaches val $0.723$ and
test $0.645$.

\begin{table}[H]
\centering
\small
\caption{Ablation of the world-model mechanisms (val/test accuracy, SHHS
160/20/20, 60 epochs, identical budget).}
\label{tab:ablation}
\begin{tabular}{lcccc}
\toprule
Version & Params & Mechanism & Val acc & Test acc \\
\midrule
Base & 196,557 & band-energy RSSM baseline & 0.714 & 0.635 \\
+Dwell sampling & 227,103 & + duration-aware sampling & 0.713 & 0.633 \\
+Dwell prior & 227,103 & + dwell survival prior \& N1 penalty & 0.721 & 0.633 \\
+Trans. calibration & 227,103 & + transition-matrix calibration & \textbf{0.723} & \textbf{0.645} \\
\bottomrule
\end{tabular}
\end{table}

\subsection{Discriminative baseline under identical conditions}
\label{sec:baseline}

Table~\ref{tab:baseline} compares the world model to the discriminative
baseline under the identical subject split and training budget. On this
single split the world model outperforms the baseline by 12.5 pp in test
accuracy and by 19.9 pp in macro-F1 ($0.600$ vs $0.401$; the baseline's N1
recall is only $0.043$). Averaged over the 3 repeated splits
(Table~\ref{tab:repsplit}) the gap is $+4.9$ pp in accuracy and $+17.5$ pp
in macro-F1: the world model is better on every fold (differences $+1.1$,
$+7.1$, $+6.4$ pp). A paired $t$-test gives $t = 2.59$ ($df=2$, $p \approx 0.12$); with
three folds the test has limited power, but the direction is consistent and
the F1 gap is large.

\begin{table}[H]
\centering
\small
\caption{Single main split (seed-consistent with Table~\ref{tab:ablation}):
world model vs discriminative baseline, same 160/20/20 split, same 60-epoch
budget, both from scratch.}
\label{tab:baseline}
\begin{tabular}{lcccc}
\toprule
Model & Val acc & Test acc & F1 & $\kappa$ \\
\midrule
Calibrated RSSM (world model) & \textbf{0.723} & \textbf{0.645} & \textbf{0.600} & \textbf{0.533} \\
Discriminative baseline & 0.624 & 0.520 & 0.401 & 0.322 \\
\bottomrule
\end{tabular}
\end{table}

\begin{table}[H]
\centering
\small
\caption{Repeated-split analysis (3 folds, mean $\pm$ 95\% CI, paired on
identical folds). The calibrated RSSM is early-stopped at 40 epochs (best validation at $\approx$
ep.~31) owing to its higher per-epoch cost, while the baseline trains for 60 epochs; both models are early-stopped on validation, and the main single-split comparison (Table~\ref{tab:baseline}) trains both for the full 60 epochs.}
\label{tab:repsplit}
\begin{tabular}{lccc}
\toprule
Model & Test acc & F1 & $\kappa$ \\
\midrule
Calibrated RSSM (world model) & $\mathbf{0.690 \pm 0.053}$ & $\mathbf{0.637 \pm 0.056}$ & $\mathbf{0.593 \pm 0.070}$ \\
Discriminative baseline & $0.641 \pm 0.042$ & $0.462 \pm 0.085$ & $0.468 \pm 0.074$ \\
\midrule
Paired diff (RSSM $-$ base) & $+0.049$, $t=2.59$ ($p\!\approx\!0.12$) & --- & --- \\
\bottomrule
\end{tabular}
\end{table}

\subsection{Dream quality: dwell distribution vs physiology}
\label{sec:dream}

\begin{figure}[!ht]
\centering
\includegraphics[width=0.9\textwidth]{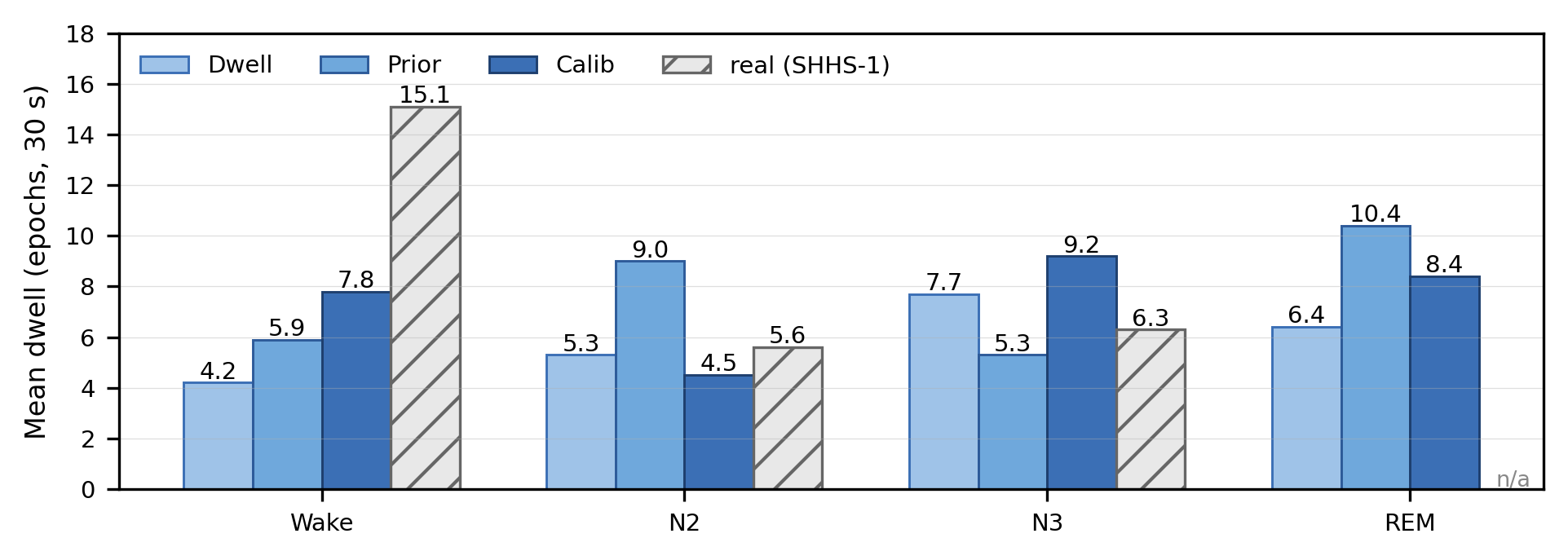}
\caption{Mean dreamed dwell per stage (Dwell/Prior/Calib) vs real \datasetA{}
statistics (20 test subjects $\times$ 80 steps, 3 dream seeds pooled). The
dwell-prior variant lengthens the N2 right tail (P90 27 vs 14 real); the
calibrated model restores the N3 frequency (N3 segments 51$\to$105; REM share
$33\%\to17\%$) at the cost of fragmented Wake runs (mean 7.8 vs 15.1 real).
Real REM dwell is not available in this subset.}
\label{fig:dwell}
\end{figure}

Because the model is generative, synthetic nights can be \emph{dreamed} by
unrolling the latent dynamics without observations. Table~\ref{tab:dream}
reports the mean dwell (in epochs) of dreamed trajectories over 20 test
subjects $\times$ 80 steps, pooled over 3 dream seeds for reproducibility,
versus the real statistics of \datasetA{} (Figure~\ref{fig:dwell}). The last
column reports the earth-mover distance (EMD) between the calibrated dreamed
and real run-length distributions; REM is not evaluable because the 80-step
real windows contain no REM runs in this subset. Run-length statistics and
survival curves are truncated at $D_{\max}=60$ epochs (30 min), a horizon
well beyond the longest physiological stage run.

Calibration corrects the N3/REM trade-off that the dwell prior introduces
(N3 segments 51$\to$105, REM share $33\%\to17\%$) but fragments Wake runs
(mean 7.8 vs 15.1 real). N1 remains over-generated (121 dreamed runs vs 15
real): N1 is rare and its EEG is close to Wake, so the model cannot separate
it reliably; this is revisited in the Discussion.

\begin{table}[H]
\centering
\small
\caption{Mean dreamed dwell (epochs) vs real \datasetA{} statistics (20
test subjects $\times$ 80 steps, 3 dream seeds pooled), plus the
earth-mover distance (EMD, epochs) between the calibrated dreamed and real
run-length distributions.}
\label{tab:dream}
\begin{tabular}{lccccc}
\toprule
Stage & Dwell & Prior & Calib & Real & EMD \\
\midrule
W & 4.2 & 5.9 & 7.8 & 15.1 & 7.3 \\
N2 & \textbf{5.3} & 9.0 & 4.5 & 5.6 & 1.2 \\
N3 & 7.7 & \textbf{5.3} & 9.2 & 6.3 & 3.0 \\
REM & 6.4 & 10.4 & 8.4 & --- & n/a \\
\bottomrule
\end{tabular}
\end{table}

\subsection{Hard constraint vs soft post-processing}
\label{sec:constraint-exp}

\begin{figure}[!ht]
\centering
\includegraphics[width=0.95\textwidth]{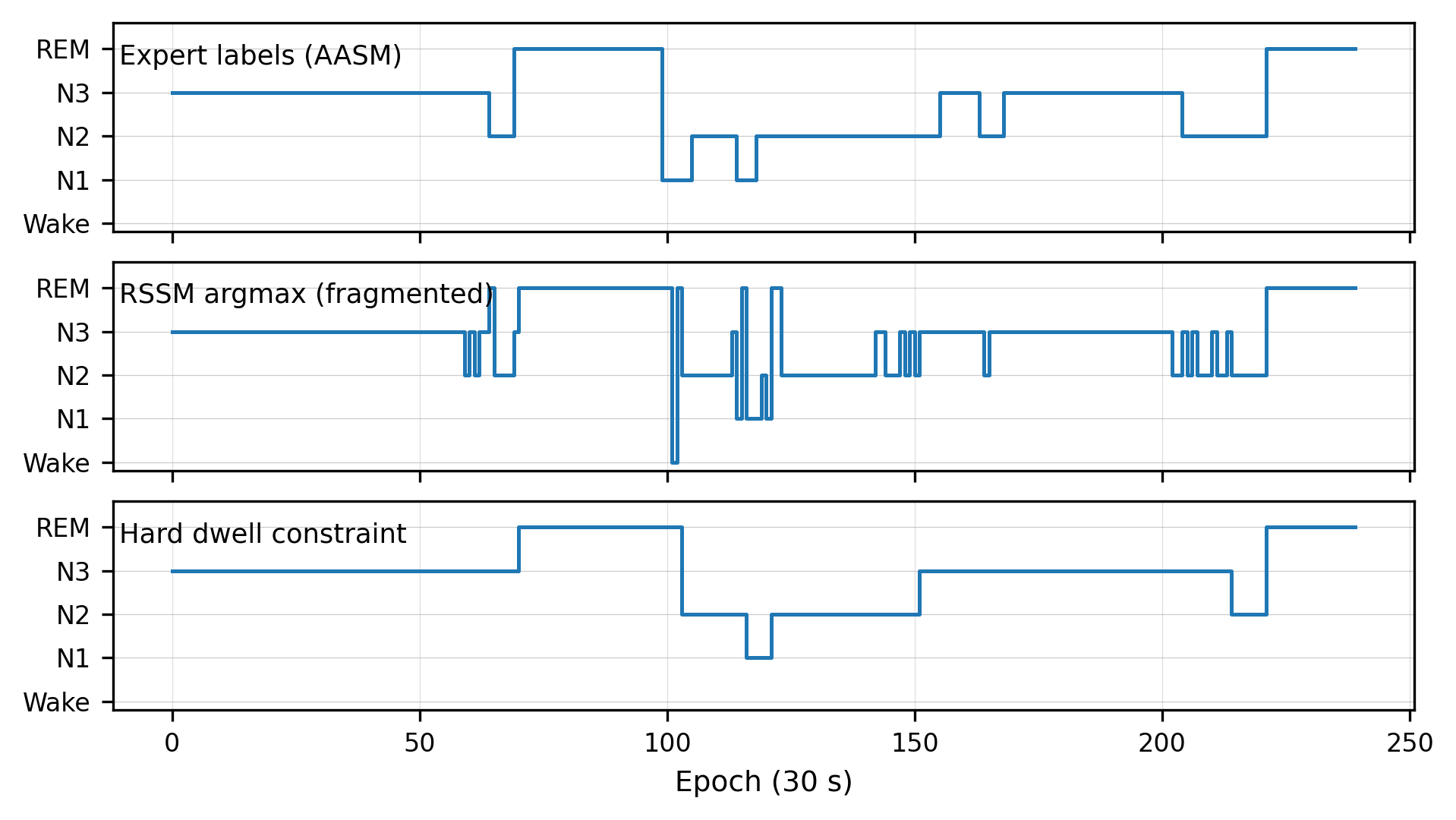}
\caption{Sleep-stage trajectory on one \datasetB{} test night (SC4001, 240
epochs): expert labels vs raw RSSM argmax (fragmented) vs the same output
after the hard dwell constraint. The constraint removes all below-$K$ runs (30 $\to$ 0 on this segment)
while preserving the macro structure.}
\label{fig:traj}
\end{figure}

The constraint layer is evaluated on \datasetB{} (32 test nights) where the
argmax of the discriminative baseline (Section~\ref{sec:baseline}) is used
as the base sequence
(the constraint is model-agnostic). With $K=\{2,2,6,4,4\}$, Table~\ref{tab:constraint}
shows that the raw argmax violates the minimum dwell on 71.8\% of runs, HSMM
Viterbi smoothing \citep{flexer2005reliable} still leaves 11.3\%, while the
run-merging constraint reduces violations to 0.1\% \emph{and} improves
accuracy ($0.734 \to 0.737$). On dreamed \datasetA{} trajectories the latent
forcing achieves 0 violations on all 5 stages. Figure~\ref{fig:traj}
illustrates the effect on one test night: the raw argmax fragments the
hypnogram with frequent 1--2 epoch runs, and the hard constraint restores a
smooth, expert-like trajectory (below-$K$: 30 $\to$ 0 runs on this segment;
the expert labels themselves contain 2).

\begin{table}[H]
\centering
\small
\caption{Below-minimum-dwell ($K$) segments on \datasetB{} 32 test nights.
Soft = HSMM Viterbi; hard = run-merging constraint.}
\label{tab:constraint}
\begin{tabular}{lccc}
\toprule
Method & Acc & Below-$K$ runs & Note \\
\midrule
Raw argmax (no constraint) & 0.734 & \textbf{71.8\%} & fragmented \\
HSMM Viterbi (soft) & 0.771 & 11.3\% & still violates \\
Hard constraint & \textbf{0.737} & \textbf{0.1\%} & accuracy up \\
\bottomrule
\end{tabular}
\end{table}

\begin{figure}[H]
\centering
\includegraphics[width=0.95\textwidth]{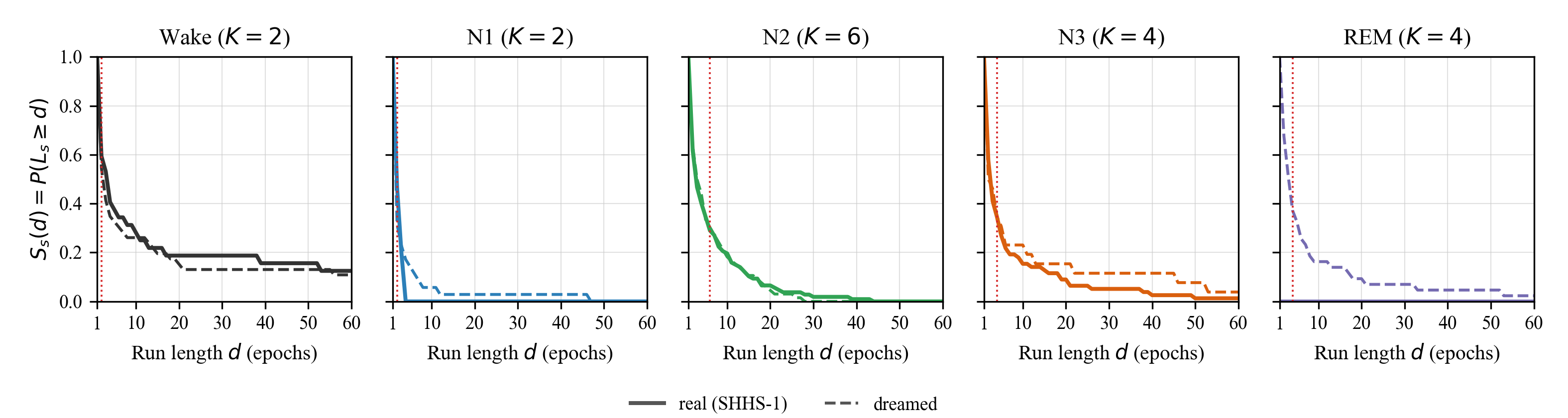}
\caption{Dwell survival curves $S_s(d)=P(L_s\ge d)$ per stage for the real
\datasetA{} test labels (solid) versus unconstrained calibrated-model dreamed trajectories
(dashed), over 20 test subjects $\times$ 80 steps, 3 dream seeds pooled. The
model matches the fast N1 decay and the N2 heavy tail; Wake runs remain
over-generated and over-fragmented, the known failure mode quantified in
Table~\ref{tab:dream}.}
\label{fig:dreamsurv}
\end{figure}

\begin{figure}[!ht]
\centering
\includegraphics[width=0.88\textwidth]{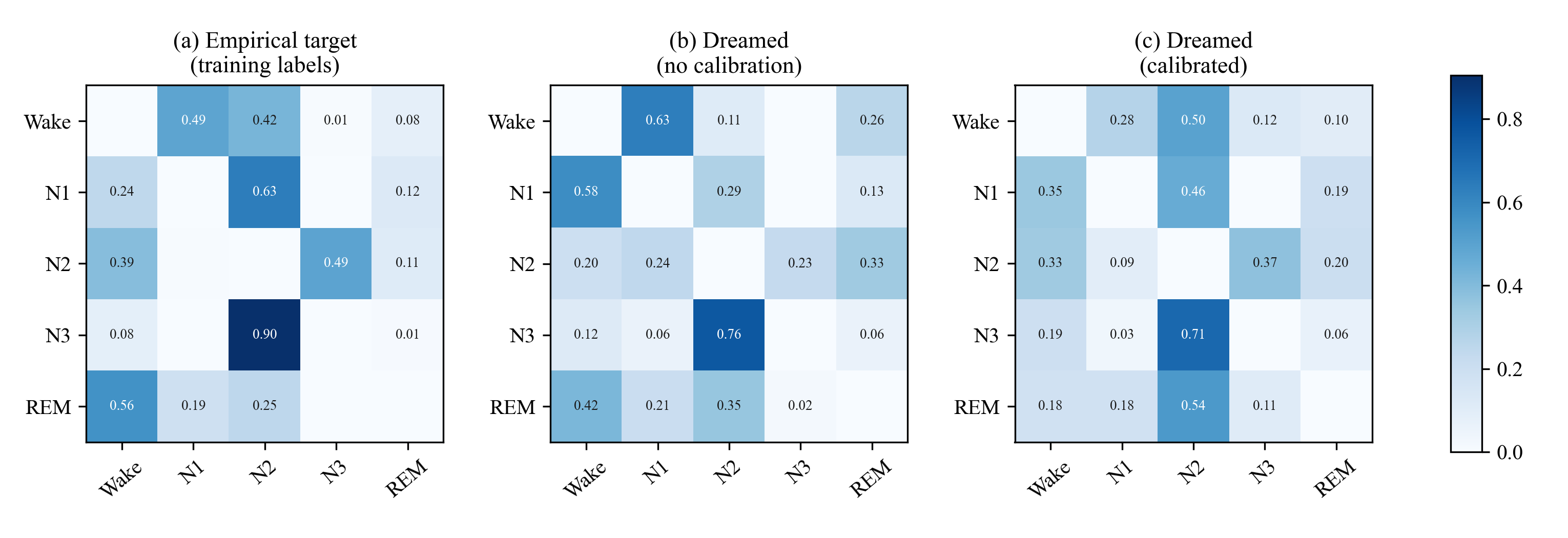}
\caption{Transition-matrix calibration. From left to right: the empirical
transition-target distribution $\hat P(t \mid s)$ estimated from the 160 SHHS-1
training labels (self-loops excluded, rows normalized); the switch frequencies
observed in the dwell-prior dreamed trajectories (no calibration); and those in the calibrated
trajectories (calibrated by Eq.~(\ref{eq:switchbias})), all pooled over 3
dream seeds. Calibration corrects the main over-transfers of the dwell-prior
variant (W$\to$N1 0.73$\to$0.43 vs 0.49 empirical; N2$\to$REM 0.42$\to$0.22
vs 0.11) at the cost of residual N3-row bias (N3$\to$N2 0.56 vs 0.90).}
\label{fig:trans}
\end{figure}

\begin{figure}[!ht]
\centering
\includegraphics[width=0.95\textwidth]{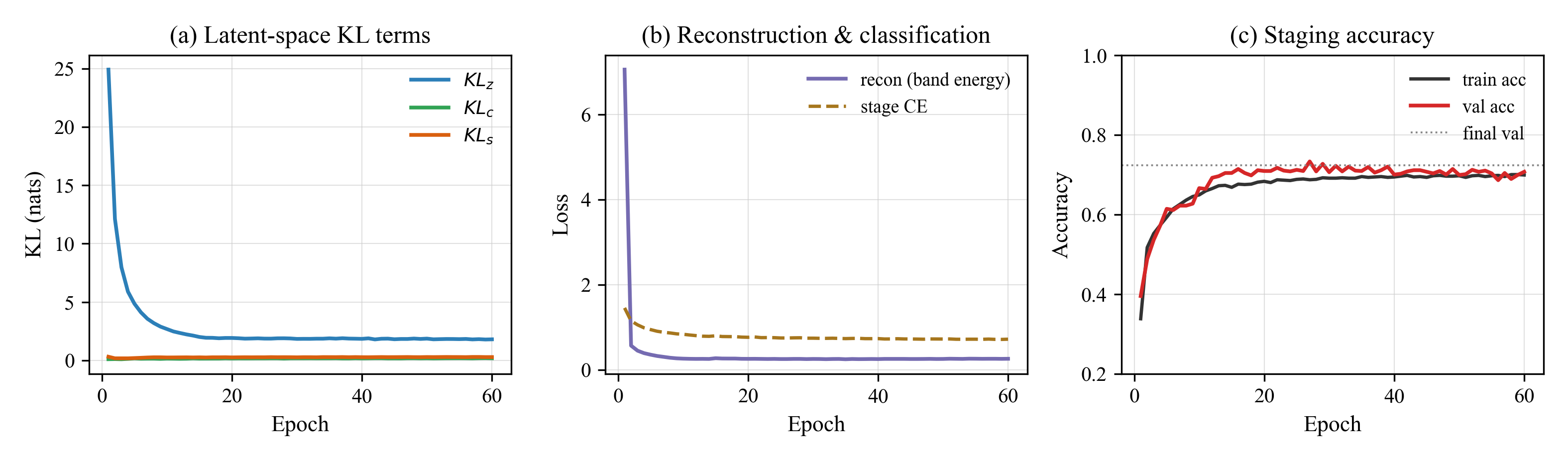}
\caption{Training health of the calibrated world model on \datasetA{} (60 epochs,
identical hyper-parameters as Table~\ref{tab:ablation}). (a) Latent-space KL
terms $\mathrm{KL}_z, \mathrm{KL}_c, \mathrm{KL}_t$ anneal and stabilize in the
healthy regime; (b) band-energy reconstruction and stage cross-entropy
converge; (c) training and validation accuracy rise steadily, matching the
reported val $0.723$.}
\label{fig:health}
\end{figure}

Figure~\ref{fig:dreamsurv} complements Table~\ref{tab:dream} with the full
dwell survival curves of the dreamed trajectories (3 dream seeds pooled): the
model reproduces the fast N1 decay and the N2 heavy tail, while Wake runs remain
over-generated and over-fragmented (Table~\ref{tab:dream}). The transition-matrix
calibration of Section~\ref{sec:transcal} is visualized in
Figure~\ref{fig:trans}: the calibrated transition frequencies move toward
the empirical target at the main over-transfer bottlenecks (W$\to$N1 from
0.73 to 0.43, versus 0.49 empirically; N2$\to$REM from 0.42 to 0.22, versus
0.11), whereas the dwell-prior trajectories over-transfer into N1 and REM; a
residual N3-row bias remains (N3$\to$N2 0.56 vs 0.90). The training health of
the 60-epoch run discussed in Section~\ref{sec:ablation} is reported in
Figure~\ref{fig:health}: the three KL terms anneal from a high start without
collapsing, reconstruction error drops to $\approx 0.25$, and validation
accuracy converges smoothly.

\vspace{-1.5em}
\subsection{Closed-loop simulation}
\label{sec:closedloop}

The online closed-loop controller is simulated on \datasetB{} 32 nights:
\looseness=-1
epochs arrive one at a time, and a causal locked scheduler commits to a stage
only after its candidate has accumulated $\ge K[\mathrm{cand}]$ epochs; once
committed, the stage is locked for $K[s]$ epochs before a switch is
considered. Table~\ref{tab:closedloop} compares the resulting policy-switch
statistics, and Algorithm~\ref{alg:lock} formalizes the scheduler. The locked
scheduler reduces switches from 306 to 41.3 per night
($-87\%$) with zero below-$K$ runs and every RL run at least 90 s (3 epochs),
at a modest accuracy cost ($0.734 \to 0.687$) that reflects the online,
causally locked setting.

\begin{table}[H]
\centering
\small
\caption{Online locked scheduling on \datasetB{} 32 nights: the locked
scheduler cuts policy switches by 87\% with zero below-$K$ runs and every RL
run $\ge$ 90 s.}
\label{tab:closedloop}
\begin{tabular}{lcccc}
\toprule
Method & Switches/night & Below-$K$ & Acc & Min RL run \\
\midrule
Raw argmax & 306 & 71.8\% & 0.734 & $<30$ s freq. \\
HSMM Viterbi & 65.6 & 11.3\% & 0.771 & --- \\
Locked scheduler & \textbf{41.3} & \textbf{0.0\%} & 0.687 & $\ge 90$ s \\
\bottomrule
\end{tabular}
\end{table}

\begin{algorithm}[H]
\caption{Causal locked scheduling}
\label{alg:lock}
\begin{algorithmic}[1]
\Require minimum dwells $K[s]$; candidate logits $\ell_t$ at each epoch $t$
\Ensure a causal stage sequence $s_1,\ldots,s_T$ whose every committed run is at least $K[s]$ epochs
\State $s \gets \mathrm{None}$; \quad $c \gets 0$; \quad $d \gets 0$ \Comment{committed stage, candidate count, locked count}
\For{$t = 1 \ldots T$}
  \State $\mathrm{cand} \gets \arg\max_s \ell_t$ \Comment{stage-wise posterior at epoch $t$}
  \If{$s = \mathrm{None}$}
    \If{$\mathrm{cand} \neq \mathrm{prev}$} \State $c \gets 1$ \Else \State $c \gets c + 1$ \EndIf
    \State $\mathrm{prev} \gets \mathrm{cand}$
    \If{$c \ge K[\mathrm{cand}]$} \State commit: $s \gets \mathrm{cand}$; $d \gets 0$ \EndIf
  \Else
    \State $d \gets d + 1$
    \If{$d \ge K[s]$ \textbf{and} $\mathrm{cand} \neq s$} \State switch: $s \gets \mathrm{cand}$; $d \gets 0$ \EndIf
  \EndIf
  \State emit $s$ \Comment{RL policy follows the locked stage}
\EndFor
\State \Return $s_1, \ldots, s_T$
\end{algorithmic}
\end{algorithm}

\vspace*{-2em}
\section{Discussion}
\label{sec:discussion}

\paragraph{Relation to published results.}
Published systems trained on the full \datasetA{} cohort reach higher
epoch-wise accuracy: SleepTransformer \citep{phan2022sleeptransformer}
reports 87.7\% accuracy and $\kappa = 0.828$ on 70\%/30\% subject splits
over 5{,}791 subjects, and other sequence models report similar
levels. The numbers reported here are not directly comparable, for three
reasons: (i) data scale---those systems train on the full cohort
(5{,}791 subjects, versus the 200 used here); (ii) training
budget---SleepTransformer is trained with an extensive
validation-driven early-stopping protocol, whereas the protocol used here
trains for 60 epochs with early stopping (a substantially smaller
budget); and (iii) model size---the RSSM used here is a
lightweight GRU-based model with 227K parameters. This paper does not aim to compete on epoch-wise accuracy; the focus is
on whether a generative world model can learn dwell structure
and guarantee stage-specific minimum dwells at decoding time without
retraining. Under identical splits and budgets the world model
still outperforms its discriminative counterpart
(Table~\ref{tab:repsplit}), and the constraint layer is model-agnostic:
it can be applied on top of any staging system, including those with
higher epoch-wise accuracy.

\paragraph{Advantage over the baseline.}
Under identical data splits the largest gap is in macro-F1: the
discriminative baseline collapses on the rare N1 class (recall 0.043),
whereas the RSSM carries temporal and generative context that helps the rare
and ambiguous classes, and the dwell mechanisms provide structural inductive
bias. This is consistent with the failure mode reported in
Section~\ref{sec:ablation}: on 47 subjects of \datasetB{} the same RSSM
training collapsed, while 200 subjects of \datasetA{} train healthily; the
base RSSM in Table~\ref{tab:ablation} trains healthily as well, so the
difference in training outcomes is driven by data scale.

\paragraph{Guarantees vs statistics.}
The hard constraint is exact at the trajectory level; the statistical
distribution of dreamed dwells is right-shifted relative to physiology (a
necessary consequence of imposing a minimum). For a closed-loop operator a guaranteed minimum is more useful than an
exactly matched distribution; on the evaluated nights run merging even
improves accuracy slightly (0.734 to 0.737, Table~\ref{tab:constraint}).

\paragraph{Limitations and generalization.}
Out-of-domain generalization remains the central limitation: the model is
trained and evaluated on in-domain SHHS recordings only, and deployment on a
new hospital, device, or unseen subject would require domain adaptation or
subject-specific personalization, which this paper does not address. N1 remains over-generated in every reported setting
(Table~\ref{tab:dream}). The repeated-split CI is limited to 3 folds by RSSM training cost
($\approx$5 h/fold); a larger number of folds would tighten the CI. Dream
quality is evaluated by dwell statistics and run-length EMD
(Table~\ref{tab:dream}); spectral FID of dreamed EEG is left
to future work.

\section{Conclusion}
\label{sec:conclusion}

The experiments show that a generative world model can stage sleep while
learning dwell structure from data, without relying on post-hoc smoothing.
Dreamed trajectories match physiological dwell statistics, a trajectory-level
hard constraint reduces below-$K$ segments from 71.8\% to 0.1\%, and an
online locked scheduler cuts policy switches by 87\% per night (every run at
least 90 s). Under identical data splits the world model also beats a
discriminative baseline by a consistent margin (test accuracy $0.690 \pm
0.053$ vs $0.641 \pm 0.042$, macro-F1 $+17.5$ pp, better on all 3 repeated
splits).

\bibliographystyle{plainnat}

\begin{thebibliography}{22}
\bibitem{perslev2019utime}
M.~Perslev, M.~H. Jensen, S.~Darkner, P.~J. Jennum, and C.~Igel.
\newblock U-time: A fully convolutional network for time series segmentation
applied to sleep staging.
\newblock In \emph{Advances in Neural Information Processing Systems
(NeurIPS)}, 2019.
\bibitem{supratak2017deepsleepnet}
A.~Supratak, H.~Dong, C.~Wu, and Y.~Guo.
\newblock Deepsleepnet: A model for automatic sleep stage scoring based on raw
single-channel EEG.
\newblock \emph{IEEE Transactions on Neural Systems and Rehabilitation
Engineering}, 25(11):1998--2008, 2017.
\bibitem{phan2022sleeptransformer}
H.~Phan, K.~Mikkelsen, O.~Y. Chen, P.~Koch, A.~Mertins, and M.~De Vos.
\newblock Sleeptransformer: Automatic sleep staging with interpretability and
uncertainty quantification.
\newblock \emph{IEEE Transactions on Biomedical Engineering},
69(8):2456--2467, 2022.
\bibitem{rabiner1989tutorial}
L.~R. Rabiner.
\newblock A tutorial on hidden Markov models and selected applications in
speech recognition.
\newblock \emph{Proceedings of the IEEE}, 77(2):257--286, 1989.
\bibitem{flexer2005reliable}
A.~Flexer, G.~Gruber, and G.~Dorffner.
\newblock A reliable probabilistic sleep stager based on a single EEG signal.
\newblock \emph{Artificial Intelligence in Medicine}, 33(3):199--207, 2005.
\bibitem{ha2018worldmodels}
D.~Ha and J.~Schmidhuber.
\newblock World models.
\newblock \emph{arXiv preprint arXiv:1803.10122}, 2018.
\bibitem{hafner2019plaNet}
D.~Hafner, T.~Lillicrap, I.~Fischer, R.~Villegas, D.~Ha, H.~Lee, and
J.~Davidson.
\newblock Learning latent dynamics for planning from pixels.
\newblock In \emph{Proceedings of the 36th International Conference on
Machine Learning (ICML)}, 2019.
\bibitem{hafner2020dreamer}
D.~Hafner, T.~Lillicrap, J.~Ba, and M.~Norouzi.
\newblock Dream to control: Learning behaviors by latent imagination.
\newblock In \emph{International Conference on Learning Representations
(ICLR)}, 2020.
\bibitem{hafner2021dreamerv2}
D.~Hafner, T.~Lillicrap, M.~Norouzi, and J.~Ba.
\newblock Mastering atari with discrete world models.
\newblock In \emph{International Conference on Learning Representations
(ICLR)}, 2021.
\bibitem{sors2018cnn}
A.~Sors, S.~Bonnet, S.~Mirek, L.~Vercueil, and J.-F. Payen.
\newblock A convolutional neural network for sleep stage scoring from raw
single-channel EEG.
\newblock \emph{Biomedical Signal Processing and Control}, 42:107--114,
2018.
\bibitem{chambon2018deep}
S.~Chambon, M.~N. Galtier, P.~J. Arnal, G.~Wainrib, and A.~Gramfort.
\newblock A deep learning architecture for temporal sleep stage classification
using multivariate and multimodal time series.
\newblock \emph{IEEE Transactions on Neural Systems and Rehabilitation Engineering},
26(4):758--769, 2018.
\bibitem{perslev2021usleep}
M.~Perslev, S.~Darkner, L.~Kempfner, M.~Nikolic, P.~J. Jennum, and C.~Igel.
\newblock U-Sleep: resilient high-frequency sleep staging.
\newblock \emph{npj Digital Medicine}, 4(1):72, 2021.
\bibitem{hafner2023dreamerv3}
D.~Hafner, J.~Pasukonis, J.~Ba, and T.~Lillicrap.
\newblock Mastering diverse domains through world models.
\newblock \emph{arXiv preprint arXiv:2301.04104}, 2023.
\bibitem{robine2023twm}
J.~Robine, M.~Hoffmann, T.~Uelwer, and S.~Harmeling.
\newblock Transformer-based world models are happy with 100k interactions.
\newblock \emph{arXiv preprint arXiv:2303.07109}, 2023.
\bibitem{vangorp2024generative}
H.~van Gorp, M.~M. van Gilst, P.~Fonseca, F.~B. van Meulen, J.~P. van Dijk,
S.~Overeem, and R.~J.~G. van Sloun.
\newblock A deep generative model for five-class sleep staging with arbitrary
sensor input.
\newblock \emph{arXiv preprint arXiv:2408.15253}, 2024.
\bibitem{yu2010hsmm}
S.-Z. Yu.
\newblock Hidden semi-Markov models.
\newblock \emph{Artificial Intelligence}, 174(2):215--243, 2010.
\bibitem{rechtschaffen1968manual}
A.~Rechtschaffen and A.~Kales.
\newblock \emph{A Manual of Standardized Terminology, Techniques and Scoring
System for Sleep Stages of Human Subjects}.
\newblock Public Health Service, U.S. Government Printing Office, 1968.
\bibitem{berry2017aasm}
R.~B. Berry, R.~Brooks, C.~E. Gamaldo, S.~M. Harding, R.~M. Lloyd, C.~L.
Marcus, and B.~V. Vaughn.
\newblock \emph{The AASM Manual for the Scoring of Sleep and Associated Events}.
\newblock American Academy of Sleep Medicine, 2.4 edition, 2017.
\bibitem{kingma2014vae}
D.~P. Kingma and M.~Welling.
\newblock Auto-encoding variational bayes.
\newblock In \emph{International Conference on Learning Representations}, 2014.
\bibitem{quan1997shhs}
S.~F. Quan, B.~V. Howard, C.~Iber, J.~P. Kiley, F.~J. Nieto, G.~T. O'Connor,
D.~M. Rapoport, S.~Redline, J.~Robbins, J.~M. Samet, et~al.
\newblock The sleep heart health study: Design, rationale, and methods.
\newblock \emph{Sleep}, 20(12):1077--1085, 1997.
\bibitem{kemp2000sleepedf}
B.~Kemp, A.~H. Zwinderman, B.~Tuk, H.~A.~C. Kamphuisen, and J.~J.~L. Obery\'e.
\newblock Analysis of a sleep-dependent neuronal feedback loop: The slow-wave
microcontinuity of the EEG.
\newblock \emph{IEEE Transactions on Biomedical Engineering},
47(9):1185--1194, 2000.
\bibitem{goldberger2000physionet}
A.~L. Goldberger, L.~A. Amaral, L.~Glass, J.~M. Hausdorff, P.~C. Ivanov,
R.~G. Mark, J.~E. Mietus, G.~B. Moody, C.-K. Peng, and H.~E. Stanley.
\newblock Physiobank, physiotoolkit, and physionet: Components of a new research
resource for complex physiologic signals.
\newblock \emph{Circulation}, 101(23):e215--e220, 2000.
\end{thebibliography}

\end{document}